# One-pot synthesis: a simple and fast method to obtain ceramic superconducting materials

*Maycon Rotta[1,2], Maycon Motta[3], Alexsander Lourenço Pessoa[1], Claudio Luiz Carvalho[1], Cesar Vanderlei Deimling[4], Paulo Noronha Lisboa-Filho[5], Wilson Aires Ortiz[3] and Rafael Zadorosny[1]*

[1]Departamento de Física e Química, Universidade Estadual Paulista (UNESP), Faculdade de Engenharia, Caixa Postal 31, 15385-000 Ilha Solteira-SP, Brazil

[2] Instituto Federal de Educação, Ciência e Tecnologia de Mato Grosso do Sul (IFMS), Campus Três Lagoas, 79641-162, Três Lagoas-MS, Brazil.

[3] Departamento de Física, Universidade Federal de São Carlos - UFSCar, 13565-905, São Carlos-SP, Brazil

[4] Departamento Acadêmico de Física, Universidade Tecnológica Federal do Paraná - UTFPR, 87301-899, Campo Mourão-PR, Brazil

[5] Departamento de Física, Universidade Estadual Paulista (UNESP), Faculdade de Ciências, Caixa Postal 473, 17033-360, Bauru-SP, Brazil

**Abstract**

The one-pot method focuses on the reduction of the number of steps or chemical reactions in the synthesis of materials, and it is very appealing in terms of sustainability. In addition to this point of view, superconductors are desired materials due to their unusual properties, such as the zero resistivity and the perfect diamagnetism. One-pot, Thus, in this work, we described the one-pot synthesis of $YBa_2Cu_3O_{7-\delta}$ superconducting ceramic. In just two steps and a few hours, a polymer composite solution was prepared, which originates a powder after burning the polymer out with pure phase and with superconducting properties better than those produced by other techniques.

## 1. Introduction

The $YBa_2Cu_3O_{7-\delta}$ (YBCO) ceramic superconductor is one of the most studied material due to its great possibility of applications[1–7]. The performance of YBCO devices depends crucially on the composition, homogeneity, and microstructure[8]. Then, the choice of its synthesis method is one of the most important steps to obtain good and low-cost material. The chemical routes such as the sol-gel ones have been successfully used since they promote the cation mixture in an atomic scale, which results in superconducting samples with a high degree of homogeneity and purity[8–10].

The Pechini Method (PM)[8,11,12] is notably an efficient method to produce compounds with various metal ions, as it is the case of the oxide superconductors, e.g., YBCO. However, such a technique requires a rigorous and exhaustive execution due to its numerous steps, reactions, and it requires a large number of reagents such as acids, bases, chelating, and polyesterification agents.

Motta *et al.*[11] have described the main reactions, as well as all the steps that occur during the YBCO synthesis via PM. Firstly, the metallic salts were dissolved separately, and just then, these solutions were mixed in a specific order. In the sequence, the chelating agent and the polyesterification agents were added. The solution was kept under stirring and heating to eliminate water, and a gel was formed. All the process counts with six distinct reactions without considering the constant monitoring of the heating and the pH control, which has to be close to 7 to avoid precipitations and to guarantee the solution polymerization. Therefore, this process can take several hours.

In this work, we explored the reducing and stabilizing properties of the polymer Poly(vinyl pyrrolidone) (PVP)[13,14] to enable the synthesis of YBCO in only one-pot[15]. The one-pot synthesis is a process that involves a series of distinct reactions in a single vessel, cutting steps off, minimizing waste generation, and saving time[15,16]. However, it still produces high-quality materials. This process is commonly used in organic chemistry[15–20], notwithstanding, it can also be used in the production of oxides materials[21,22], nanocrystals[23,24], and so on[18,25,26]. As a result, YBCO samples were produced simply, with a reduced number of reagents, time, and in only two steps. The X-ray diffractometry (XRD) showed the formation of a pure phase and the magnetic characterizations confirmed the achievement of a high-quality superconductor, with the critical temperature ($T_c$) of 92.7 K, synthesized in a simple way and, most importantly, with equal or better properties than the materials produced by chemical routes such as the PM[8,11].

## 2. Experimental procedures
### 2.1. Materials

The precursor solutions were produced by using the following reagents: poly(vinyl pyrrolidone) (PVP, Mw = 360,000), yttrium acetate hydrate [$Y(CH_3CO_2)_3 \cdot xH_2O$] (99.9%), barium acetate [$Ba(CH_3CO_2)_2$] (99%), copper acetate monohydrate [$Cu(CH_3CO_2)_2 \cdot H_2O$] (99%), methanol, acetic acid, and propionic acid. All the chemicals were purchased from Sigma Aldrich.

### 2.2. One-pot synthesis

To obtain 2g of YBCO in an exact stoichiometry, i.e., Y:Ba:Cu = 1:2:3, 0.7995(1) g of yttrium acetate, 1.5490(3) g of barium acetate, and 1.8162(1) g of copper acetate, totaling 4.1647(4) g of solid reagents, were sequentially added in a single pot. Then, under magnetic stirring, 3.46 ml of acetic acid, 5.2 ml of propionic acid, and 8.65 ml of methanol were added, totaling 17.31 ml of solvents. Keeping the beaker hermetically closed, to prevent early evaporation of the reducing agents, 1.0823(7) g of PVP was added. After 1 h of stirring, another 0.5 g of PVP was added, and finally, after other 30 min of stirring, the remaining 0.5 g of PVP was dissolved, in a total of 2.0823(7) g of PVP. In the sequence, the beaker was opened, and the solution was left in stirring for two more hours. The entire process was conducted at room temperature and under constant magnetic stirring, resulting in a dark green solution without precipitates with high viscosity and pH = 4.0. Briefly, the solvent was composed of 50% methanol, 30% propionic acid, and 20% acetic acid in volume. The concentration of PVP relative to the solvent was 12% (w/v), and the ratio of PVP (in terms of the repeating unit/monomers) to metallic cations ($M^{n+}$) present in the solution was PVP:$M^{n+}$ = 1.04:1.

## 2.3. Heat Treatment

The heat treatment of the gel occurred in three steps intercalated by grinding in an agate mortar. First of all, the gel was treated at 200 °C/3h at a controlled heating rate of 3 °C/min. In the second step, the powder was calcined at 600 ºC/3h at a heating and cooling rate of 2 ºC/min to eliminate PVP and other organic compounds. In the third step, a powder was obtained and divided into four batches, which were separately heated again at 600 °C under heating and cooling rate of 3 °C/min. Then the samples were calcined at different temperatures for 30 minutes in a tubular furnace, as summarized in Table 1.

## 2.4. Pechini sample

A reference sample was synthesized following the Pechini route as described in Ref. 11. In such work, Motta and coworkers studied the influence of different chelating agents, i.e., citric acid, Ethylenediaminetetraacetic acid (EDTA), and tartaric acid (TA) on the morphological and superconducting properties of YBCO samples. The TA sample presented the best superconducting signal, and then we used it as a reference sample. Such sample was heat treated at 900 ºC following a heat and cooling rate of 10 ºC/min. In Table 1 we show a summary of the samples synthesized in this work, the heating and cooling rates, and the average crystallite sizes obtained by applying the Debye-Scherrer equation[31] in the most intense peak of the XRD data.

Table 1 – Details of the samples prepared, comparing the calcination temperature, the heating and cooling rates, and the average crystallite size.

| Samples | Calcination temperature (ºC) | Heating and cooling rate (ºC/min ) | Average Crystallite size (nm) |
|---|---|---|---|
| **850A** | 850 | 2 | 30,7 ± 1,5 |
| **900A** | 900 | 2 | 24,9 ±1,3 |
| **900B** | 900 | 1 | 25,5 ± 1,8 |
| **900TA** | 900 | 10 | 27,1 ± 1,6 |

## 2.5. Characterization

Scanning electron microscopy (SEM) analysis was performed using an EVO LS15 Zeiss operated at 20 kV. Thermogravimetric measurements were performed using a TA Instruments model Q600 at temperatures between 25 and 1000 °C with a heating rate of 10 °C/min. Analyses were performed in a nitrogen atmosphere with a flow rate of 100 mL/min. XRD was performed using a Shimadzu XDR-6000 diffractometer with CuKα radiation (wavelength: 1.5418 Å). The displacement ranged from $2\theta = 5°$ to $60°$ at a scan rate of 0.02°/min. The magnetic measurements were carried out in a Quantum Design MPMS-5S magnetometer (SQUID).

## 3. Results and discussion

Regarding the solvents chosen, the high solubility of the selected metal salts (cations sources) at room temperature was the main factor responsible for reducing the number of reagents used and for obtaining a solution without precipitates. Sources of cations with low solubility may result in non-homogeneous phases due to precipitation during precursor solution synthesis. PVP is a very important component in this synthesis because it acts as a stabilizer and as a reducer. Its final hydroxyl (-OH) group is considered an ideal reducer for the synthesis of metallic nanostructures. According to Xiong et al. [12], greater control of the reduction kinetics can be obtained by varying the molar ratio between the PVP monomers and the metal salts. Additionally, the adsorption of PVP on the surfaces of metallic nanostructures can be facilitated by the oxygen and nitrogen atoms of the pyrrolidone unit, making it a good stabilizer, protecting and avoiding the agglomeration of the desired material[13,14].

In this synthesis, there was no need for pH or temperature control neither chelating agents nor polymerization elements were used, whose presence should add many impurities, as observed in the PM[8,11,12]. Proportionally comparing the production between the weight of the ceramic powders and the amount of polymer used, in the one-pot method presented here, we have a production rate of 2.0 instead of 0.15 from the PM[27].

The main panel of Figure 1 simultaneously shows the graphs of the thermal analysis (TGA/DTA) for the precursor solution (before the heat treatments and after drying the solution overnight at 100°C). The mass loss is about 33% between 80°C and 260°C, which is associated with dehydration and volatile elimination. In the range of 260°C to 600°C, the successive endothermic peaks correspond to the complete degradation of the polymer and the decomposition of the organic materials[1,28,29]. In this temperature range, the resulting mass loss is 34%. The total mass loss from room temperature at 925°C was approximately 74%, confirming the low presence of impurities after calcination. In the inset of Figure 1, the thermal analysis of the powder after calcination at 600°C/3h is shown, where successive exothermic peaks above 800°C can be observed. These peaks were suppressed by a large amount of PVP and other organic materials present in the gel solution shown in the main panel of Figure 1. The endothermic peaks and the final mass loss between 750°C and 860°C are indicative of the beginning of YBCO crystallization[28–30], represented by an exothermic peak starting at 860°C. These results were used to indicate the appropriate choice of heat treatment temperatures to obtain the YBCO phase.

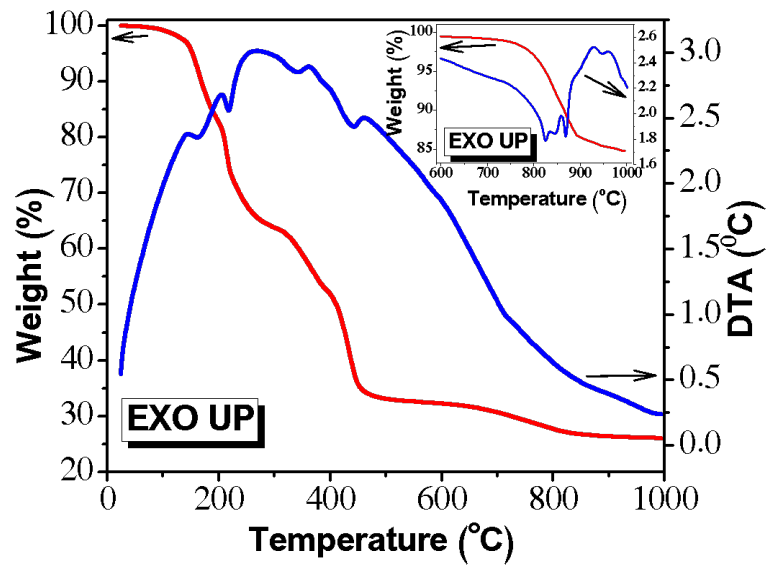

Figure 1. (Color online) Simultaneous TGA/DTA thermal analysis of the as-produced samples. The main panel shows the curves of the gel solution after dried overnight at 100°C, and in the inset, the powder produced after the second calcination step is presented, treated at 600°C/3h. The peaks above 800°C indicate the beginning of the YBCO phase crystallization.

Considering the SEM micrographs shown in Figure 2, it is possible to observe that the use of higher calcination temperatures and lower heating and cooling rates increase the coalescence of the particles, forming aggregates. Additionally, such a variable highly influenced the calcination process since it is a diffusion process. Then, the right choice of the calcination temperature and heating/cooling rates are of crucial importance to obtain high-quality materials. In Figure 2 (a), it can be seen that aggregates of small grains form the powder of sample 850A. However, by fixing the heat and cooling rates at 2°C/min and increasing the calcination temperature to 900°C (sample 900A, Figure 2 (b)), it can be seen that the aggregates are now composed of larger particles, indicating the coalescence among the prior smaller grains. Figure 2 (c) shows the morphology of sample 900B, in which we used heating and cooling rates of 1°C/min (calcination at 900°C). It is noticed that the lower heating/cooling rates promote the coalescence of the particles. As a consequence, the aggregates consist of larger and denser particles. In Figure 2 (d) the sample 900TA is shown. The powder obtained has smaller grains than those produced by the one-pot route. However, it also presented aggregates.

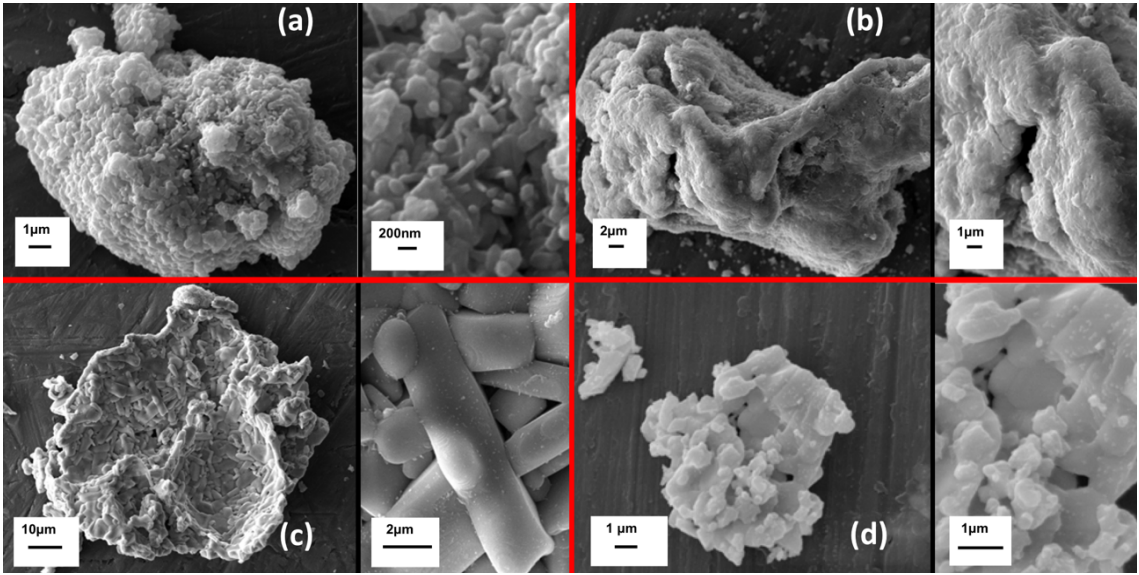

Figure 2 – SEM micrographs showing the aggregate morphology and the increasing coalescence of the particles as the thermal treatment parameters are changing for samples: (a) 850A, (b) 900A, and (c) 900B. In panel (d) sample 900TA is shown, which presents smaller grains with several aggregates.

Figure 3 shows the X-ray diffractograms of the samples. In (a), sample 850A, there is a great number of secondary phases, such as CuO, $Y_2O_3$, and $BaCO_3$. In (b), sample 900A, the $YBa_2Cu_3O_{7-\delta}$ phase is predominant, but secondary phases are still present. In curve (c) and Figure 9 of Ref. 11, for the samples 900B and 900TA, respectively, we observed the formation of pure YBCO phase with no secondary phases. A comparison between the diffractograms of samples 900A and 900B shows that the heating and cooling rates of 1°C/min provided samples with a low number of secondary phases.

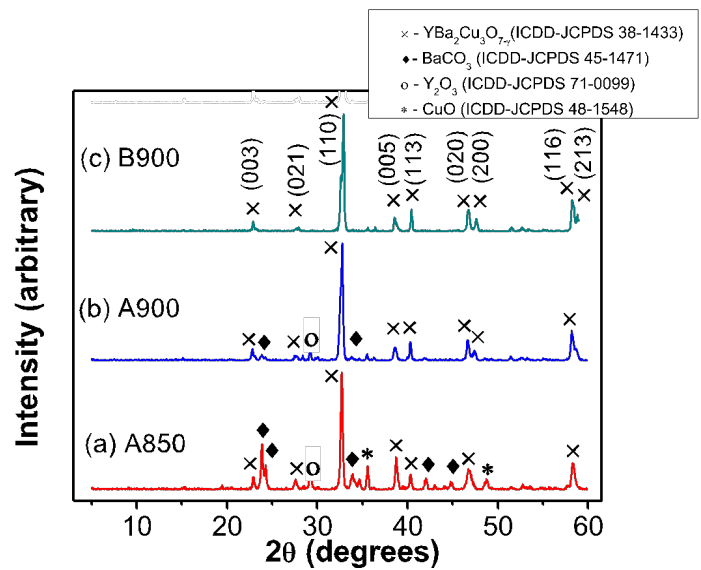

Figure 3 - XRD diffractograms for the YBCO superconducting samples (a) 850A, (b) 900A, and (c) 900B. The XRD of sample 900TA is shown in Figure 9 of Ref. 11.

The average size of the crystallite size was estimated by using the Debye-Scherrer equation[31] in the peak at 2θ= 32.9° of the diffractograms presented in Table 1. It is possible to observe that, by increasing the calcination temperature, the average crystallite size decreases from 31 nm to ~25 nm for the samples 850A and 900A and 900B, respectively. However, the sample 900TA presented an average crystallite size of ~27 nm.

Figure 4 shows the AC magnetic susceptibility as a function of the temperature for the samples 900A, 900B, and 900TA. The data were normalized by the mass of the samples. It can be noticed that the magnetic behavior of the samples is clearly influenced by the calcination temperature and by the heating and cooling rates. Higher temperatures and lower rates showed larger diamagnetic signals. The sample 850A did not present a superconducting signal, as expected due to the lack of superconducting phase as presented by the X-ray data. When comparing the samples 900A and 900B, the critical temperatures are 91.7 K and 92.7 K, respectively. It is possible to see that lower heating and cooling rates improve $T_c$. On the other hand, the diamagnetic response of the samples suggests that the rate of 1°C/min used in the production of sample 900B increases the homogeneity in the grains, as well as in the region among them (intergrain).

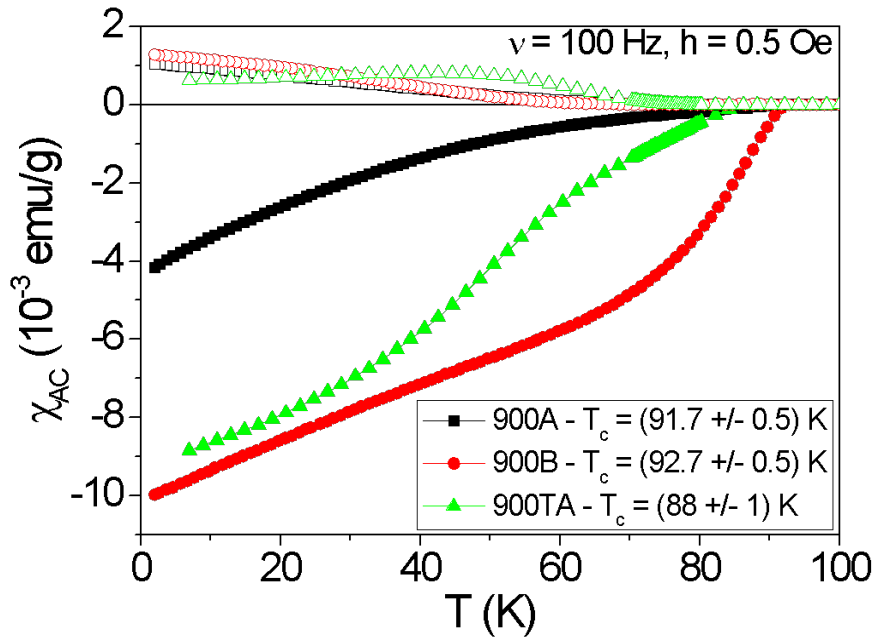

Figure 4 – Real component of the AC magnetic susceptibility versus temperature for samples 900A, 900B, and 925B.

## 4. Conclusions

The One-pot synthesis was successfully used to obtain YBCO superconducting samples with good quality and a critical temperature around 92 K. This synthesis route proved to be an efficient, simple, fast, and inexpensive alternative to produce ceramic materials when compared to other routes. The high stability of the precursor solution is directly associated with the interaction between PVP and the cations, which are sterically captured by the polymer chain. The application of the One-Pot synthesis and productivity are directly related to the right choice of precursor salts since high solubility improves the efficiency of the process. Low temperatures and high heating and cooling rates during the heat treatment hinder the coalescence of grains, resulting in small aggregates and preventing the desired YBCO phase formation. Therefore, superconducting samples were composed of aggregates with larger and denser particles with a higher diamagnetic response.

**Acknowledgments**

We thank the Brazilian Agencies São Paulo Research Foundation, FAPESP, grant 2016/12390-6, Coordenação de Aperfeiçoamento de Pessoal de Nível Superior - Brasil (CAPES) - Finance Code 001, and National Council of Scientific and Technological Development (CNPq, grant 302564/2018-7).
MR and ALP produced the one-pot samples under supervision of CLC and RZ. MM, CVD produced the Pechini samples and carried out the magnetic measurements, both under the supervision of PNLF and WAO. All authors contributed to the analysis of the data, and the writing of the manuscript. MR, MM, WAO, and RZ conceptualized the final work. RZ was responsible for the funding acquisition, and project administration.